\documentclass[aps,10pt, pre,reprint,superscriptaddress,showpacs,longbibliography]{revtex4-1}

\usepackage{epsf}
\usepackage{graphicx}
\usepackage{subfigure}
\usepackage{amsmath}
\usepackage{float}
\begin{document}
\title{Lifetimes of ultralong-range strontium Rydberg molecules in a dense BEC}
\author{J.~D.~Whalen}
\author{F.~Camargo}
\author{R.~Ding}
\author{T.~C.~Killian}
\author{F.~B.~Dunning}
\affiliation{Department of Physics and Astronomy, Rice University, Houston, TX  77005, USA}
\author{J.~P\'erez-R\'{\i}os}
\affiliation{Department of Physics and Astronomy, Purdue University, West Lafayette, IN  47907, USA}
\affiliation{School of Materials Sciences and Technology, Universidad del Turabo, Gurabo, PR 00778, USA}
\author{S.~Yoshida}
\author{J.~Burgd\"{o}rfer}
\affiliation{Institute for Theoretical Physics, Vienna University of Technology, Vienna, Austria, EU}

\begin{abstract}
The lifetimes and decay channels of ultralong-range Rydberg molecules created in a dense BEC are examined by monitoring the time evolution of the Rydberg population using field ionization.  Studies of molecules with values of principal quantum number, $n$, in the range $n=49$ to $n=72$ that contain tens to hundreds of ground state atoms within the Rydberg electron orbit show that their presence leads to marked changes in the field ionization characteristics. The Rydberg molecules have lifetimes of $\sim1-5\,\mu$s, their destruction being attributed to two main processes: formation of Sr$^+_2$ ions through associative ionization, and dissociation induced through $L$-changing collisions.  The observed loss rates are consistent with a reaction model that emphasizes the interaction between the Rydberg core ion and its nearest neighbor ground-state atom.  The measured lifetimes place strict limits on the time scales over which studies involving Rydberg species in cold, dense atomic gases can be undertaken and limit the coherence times for such measurements.
\end{abstract}

\maketitle

\section{Introduction}
\label{S:into}
Scattering of the excited electron in a Rydberg atom from neighboring ground-state atoms can lead to formation of weakly-bound ultralong-range Rydberg molecules that comprise a Rydberg atom and one, or more, ground-state atoms.  The interaction between the excited Rydberg electron and a ground-state atom can be described using a Fermi pseudopotential and results in a molecular potential that can bind multiple vibrational levels with energies that depend on the value of the principal quantum number, $n$, and that range from several tens of megahertz for $n\sim30$ to a few hundreds of kilohertz for $n\sim72$.  The existence of such a novel bond was first predicted theoretically \cite{PhysRevLett.85.2458} but has now been observed using a variety of species including rubidium, cesium, and strontium.  Initial experiments focused on creation of dimer molecules comprising a ground-state rubidium atom bound to a spherically-symmetric Rb(ns) Rydberg atom \cite{bbn09}.  Measurements have since been  extended to include other Rydberg species, anisotropic P and D Rydberg states and, using Cs(ns) Rydberg atoms, to the creation of so-called trilobite states with very large permanent electric dipole moments \cite{Li1110,PhysRevLett.109.173202,PhysRevA.92.031403,PhysRevLett.111.053001,PhysRevLett.114.133201,PhysRevLett.112.143008,PhysRevLett.112.163201,Booth99,PhysRevLett.115.193201,elp17}.  Detailed spectroscopic studies have revealed the formation of molecules comprising one Rydberg atom and up to four bound ground-state atoms \cite{gkb14}.  Measurements of molecule formation in rubidium have also been extended to BECs whose peak densities can approach $\sim10^{15}$~cm$^{-3}$ \cite{dky16,bbn11}.  For such densities, even for moderate values of $n$, $n\gtrsim50$, the electron orbit can enclose tens to hundreds of ground-state atoms.

Recent work has suggested that Rydberg molecules might be used to probe the
properties of cold dense gases and examine collective phenomena such as the
creation of polarons in quantum degenerate gases \cite{csw17}, and
to image the Rydberg electron wave function
\cite{1367-2630-17-5-053046,nte16}.  Such studies, however, require that the molecular
lifetimes be sufficient to allow interactions time to produce measurable
effects.  This has stimulated interest in the lifetimes of Rydberg molecules
and in the mechanisms that lead to their destruction.

Initial studies of Rydberg molecule lifetimes focused on Rb(35s)-Rb dimers created in a cold
gas \cite{bbn11}.  This work showed that the lifetime of the molecules was
significantly shorter than that of the parent Rydberg atoms and decreased with increasing vibrational excitation.  This behavior was attributed to the presence of a strong p-wave shape resonance in low-energy electron scattering from rubidium.  A significant increase in the molecular loss rate was also observed with increasing atom density, the cross section for loss being comparable to the geometric size of the molecule.  In contrast, subsequent studies using Sr(5s38s $^{3}$S$_{1}$)-Sr dimers  revealed that the lifetimes of the low-lying dimer vibrational states were very similar to that of the parent atomic Rydberg state \cite{PhysRevA.93.022702}. In addition, these measurements showed that the strontium molecular lifetimes were much less sensitive to the atom density in the trap than is the case for rubidium. Lifetime measurements in rubidium have recently been extended to BECs to explore the chemical reactions responsible for Rydberg-molecule destruction \cite{PhysRevX.6.031020}.

In the present work we explore the lifetimes of Rydberg molecules with $n=49, 60,$ and 72 created in a
BEC of $^{84}$Sr with peak density $\sim4\times10^{14}$~cm$^{-3}$.
 Measurements reveal sizable loss rates,
$\sim 2-8\times10^{5}$~s$^{-1}$. 
At the low temperatures characteristic of a BEC, the Langevin cross section for ion-atom
collisions~\cite{PhysRevLett.102.223201,PhysRevA.62.012709} is large and is comparable to the size of a Rydberg molecule with $n\sim50$.
Thus, when a sizable number of 
ground state atoms are present within the Rydberg molecule, the interaction
between the Rydberg core ion and a neighboring ground-state atom can
become dominant over that between the Rydberg electron and the ground-state atom. However, inelastic scattering of the Rydberg electron during a core-ion - ground-state atom interaction can result in molecule destruction through associative ionization and through dissociation induced by $L$-changing reactions in which the Rydberg electron transitions to a lower-lying, higher-$L$ state \cite{PhysRevA.20.1890,PhysRevA.21.819}. The present measurements show that the molecular loss rates associated with each of these reaction channels, $\sim2-4\times 10^{5}$~s$^{-1}$, are much larger than those associated with radiative decay of the molecules.  The collisional loss processes, and their associated rates, are similar to those seen in the earlier studies using rubidium  even though strontium  has no p-wave electron scattering resonance\cite{PhysRevX.6.031020}.  The present measurements are consistent with the predictions of a classical model of the reaction dynamics which emphasizes the role of the interaction between the Rydberg core ion and nearest ground-state atom.

\section{Experimental method}
The techniques used to cool and trap strontium atoms have been described in detail elsewhere \cite{ssk14,PhysRevLett.103.200402}.  Briefly, strontium atoms are initially cooled in a ``blue'' magneto-optical trap (MOT) operating on the 5s$^{2}$ $^{1}$S$_{0}$-5s5p $^{1}$P$_{1}$ transition at 461~nm, whereupon the atom temperature is further reduced using a ``red'' MOT utilizing the narrow 5s$^{2}$ $^{1}$S$_{0}\rightarrow$5s5p $^{3}$P$_{1}$ intercombination line at 689~nm.  The atoms are then loaded into an approximately spherically-symmetric optical dipole trap (ODT) formed by two crossed 1.06$\mu$m laser beams with $\sim 60\,\mu$m waists where they are subject to evaporative cooling to create a BEC.  The peak trap density was determined from measurements of the total atom number and trap oscillation frequencies. Typically $\sim4\times 10^5$ atoms are trapped with a peak density of $\sim 4 \times 10^{14}$~cm$^{-3}$.  Considering the radii of $n=49$ and $n=72$ atoms, given by $\sim2(n-\delta)^{2}$ (unless otherwise noted atomic units are used throughout), where $\delta=3.371$ is the quantum defect for $^{3}$S$_{1}$ Rydberg states, at the peak density  $\sim15$ and $\sim170$ atoms are present within the Rydberg electron cloud, respectively.  The temperature of the atoms is estimated to be $\sim150$ nK and the condensate fraction $\eta$ is $\sim75\%$ \cite{csw17}.

The stray electric field in the ODT was estimated from spectroscopic measurements which showed that well-defined Rydberg features could be observed for values of $n \lesssim 130$. Given that the field at which states in neighboring $n$ manifolds first cross is $1/3n^5$, this suggests that the stray fields are well below this limit, i.e., well below 45 mV cm$^{-1}$.

The Rydberg atoms and molecules are created by two-photon excitation via the
intermediate 5s5p $^{3}$P$_{1}$ state which requires radiation at 689 and
319~nm.  The 689~nm laser for the first step is tuned 80~MHz to the blue of
the intermediate state to avoid scattering from the atomic state.  The 689 and
319~nm lasers cross at right angles and have linear and circular polarizations
respectively, resulting in the production of $^{3}$S${_1}$ states with
$M=\pm$1. The 319~nm laser is tuned to the atomic or molecular state of
interest.  The excitation lasers are chopped to form a periodic train of
optical pulses with a pulse repetition frequency of $\sim4$~kHz and a pulse
duration of $\sim2 \mu$s.  The ODT was turned off during excitation to
eliminate AC Stark shifts.  The number of Rydberg atoms or molecules created during a single laser pulse was kept small, $<0.1$, to minimize any possible effects associated with blockade.

The number of surviving Rydberg atoms (molecules) is measured as a function of
time delay, $t_{D}$, following the laser pulse by application of an electric
field ramp that ionizes the Rydberg atoms (molecules).  The applied fields
have rise-times of $\sim1 \mu$s and peak values sufficient to ionize all
Rydberg states. The number of electrons liberated is determined as a function
of time during the ramp by directing them to a microchannel plate (MCP)
detector whose output pulses are fed to a multichannel scalar (MCS) that has a
minimum bin width of 100~ns.  The electric field at which ionization occurs is
determined from the electron arrival time at the MCP and the time dependence
of the electric field, which was calculated from the measured time dependence
of the voltage pulses applied to a series of electrodes positioned around the
ODT.  The uncertainty in the calibration of the applied fields experienced by
the atoms (molecules) in the ODT is estimated to be $\lesssim10\%$.  The resolution with which the ionizing field can be determined is governed by the bin width, i.e., time resolution, of the MCS.

\section{State-selective field ionization spectra}
\begin{figure}
\includegraphics[scale=0.8]{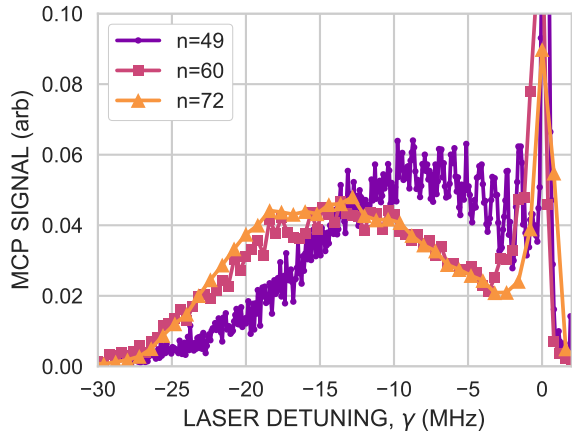}
\caption{Excitation spectra recorded for 49s, 60s, and 72s triplet Rydberg
  states as a function of detuning from the atomic line. The spectra
  for different $n$ are normalized to have equal areas.}
\label{fig:Fig1}
\end{figure}

Excitation spectra recorded for 49s, 60s, and 72s Rydberg states are shown in  Fig.~\ref{fig:Fig1} expressed as a function of detuning from the atomic excitation line.  The sharp peak at zero detuning results from the excitation of the thermal, non-condensed atoms present in the trap.  The majority of these atoms surround the BEC and their density is relatively low.  
Each spectrum also includes a broad asymmetric feature that extends to the red of the atomic line which, as earlier studies have shown \cite{csw17}, is associated with many-body interactions.  The structure evident in the 49s spectrum at small detunings results from the population of (well-resolved) bound states of dimer, trimer, tetramer\ldots~molecules having one, two, three\ldots~bound ground-state atoms.  The binding energy of the lowest-lying dimer $\upsilon=0$ vibrational state scales as 1/$(n-\delta)^{6}$ and for $n=49$ amounts to $\sim1.7$ MHz (the $\upsilon=1$ an $\upsilon=2$ dimer states have significantly smaller binding energies~\cite{PhysRevA.93.022702}.)  The binding energies for the trimer, tetramer\ldots~ Rydberg molecules are higher and  are given
approximately by distributing the bound atoms among the available dimer
levels.  For values of $n\gtrsim60$ the linewidth of the 319~nm laser,
$\sim400$ kHz,  is such that individual molecular levels can no longer be resolved.  For a given value of $n$, i.e., a given atomic volume, the number of ground-state atoms within the molecule increases steadily with detuning which, in turn, requires excitation in regions of increased local density, $\rho$.  For large ground-state atom densities,  the detuning $\Delta E$ can be approximately related to the local density, $\rho$,~\cite{csw17} using the mean field expression
 
 \begin{equation}\label{eq:large detunings}
 \Delta E=2\pi A_{s}(k)\rho 
 \end{equation}
 where $A_{s}(k)$ is the effective momentum-dependent s-wave scattering length derived in \cite{polaronPRA}.

\begin{figure}
\includegraphics[scale = 0.77]{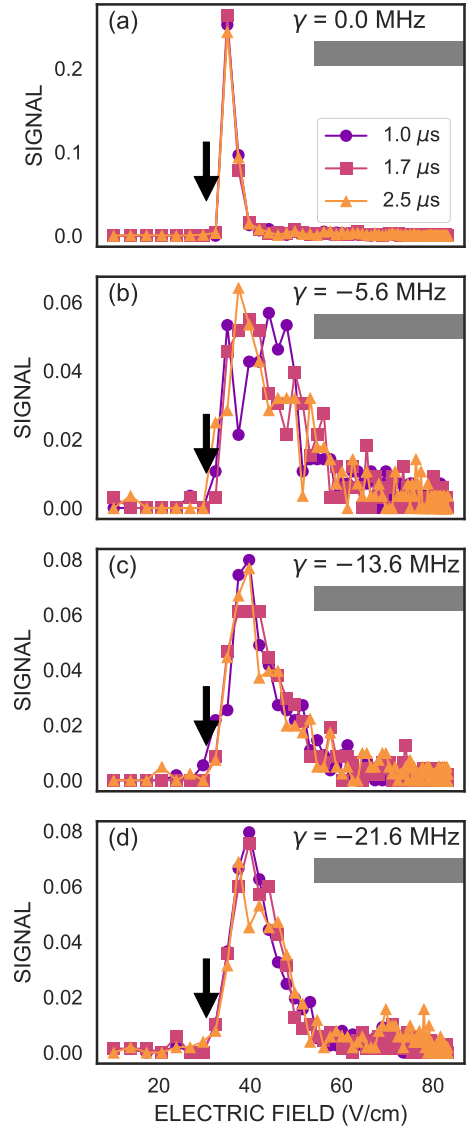}
\caption{SFI spectra recorded at $n=60$ for the delay times $t_{D}$ indicated.   (a) Spectra obtained for the parent atomic state with the laser tuned to atomic resonance; (b), (c), and (d) spectra recorded for Rydberg molecules with the detunings, $\gamma$, indicated. The arrows show the threshold field for adiabatic ionization ($1/16(n-\delta)^4$), the gray bars the fields, which extend to $\sim120\,\textup{V}\,\textup{cm}^{-1}$, over which purely diabatic ionization is expected. The various profiles in each figure are normalized to have equal area.}
\label{fig:SFI_spectra}
\end{figure}

\begin{figure}
\includegraphics[scale = 0.45]{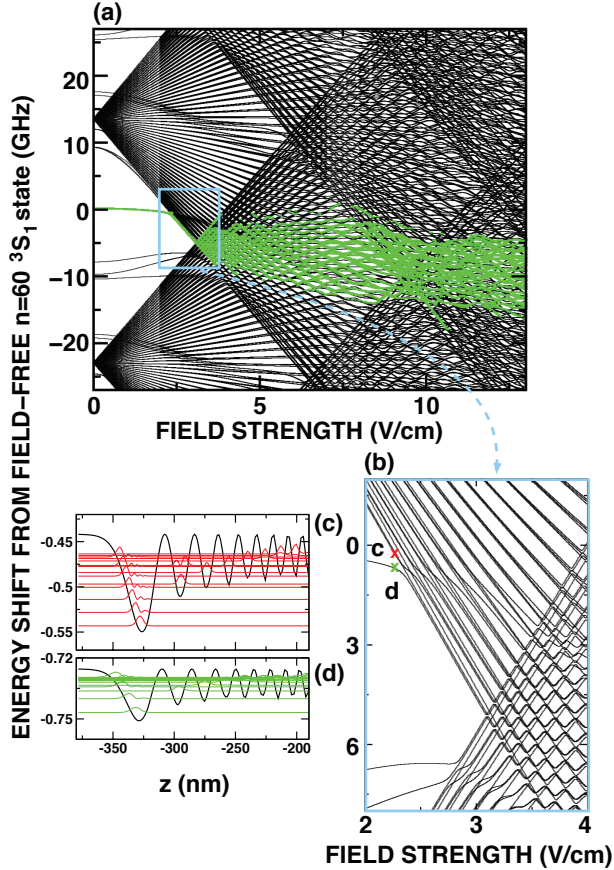}
\caption{(a) Calculated Stark energy level structure for $M_{J}=1$ strontium
  atoms in the vicinity of the $n=60$ $^{3}$S$_{1}$ state.  The green lines
  show the states with more than 1\% occupation probability when ionizing the $^{3}$S$_{1}$ state using a field 
  with a slew rate of 43 V cm$^{-1}$
  $\mu$s$^{-1}$.  Panel (b) shows a portion of this structure on an expanded
  scale.  Panels (c) and (d) show, respectively, the calculated molecular
  potentials, V($\vec{R}$), for a ``dimer'' molecule, formed from (c) the most
  red-shifted Stark state and (d) the parent $n=60$ $^3$S$_1$ state in a field
  of 2.3\,V\,cm$^{-1}$.  The molecular wave functions,
  $\vert R_g\psi_\lambda (\vec{R}_g)\vert^{2}$, for the lowest 
  lying rovibronic levels $(\lambda = 1, 2, \cdots, 16)$ 
  are also shown (see text). (c) and (d) show cross sections at 
  $\vec{R}_g =(0,0,z)$.}
\label{fig:Stark_diagram} 
\end{figure}


\begin{figure}
\includegraphics[scale = 0.4]{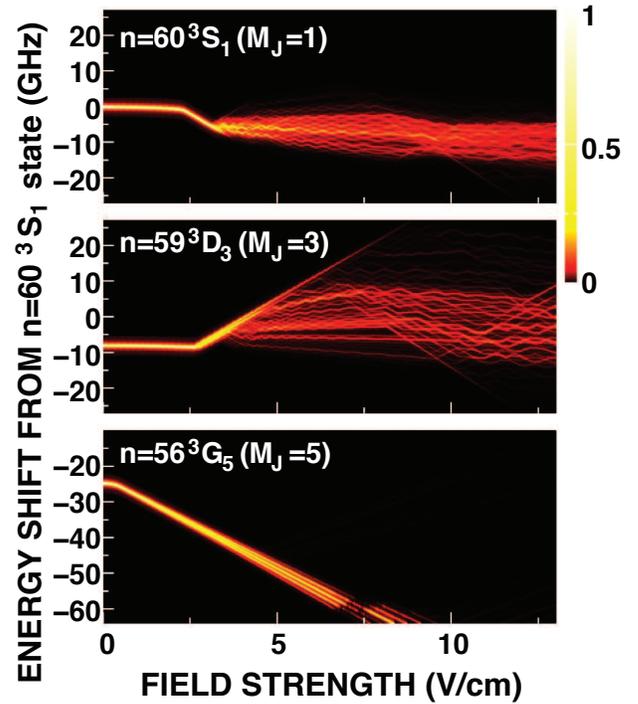}
\caption{Calculated time evolution of occupation probability during ionization of the states
  indicated using a field with a slew rate of 43 V cm$^{-1}$ $\mu$s$^{-1}$.
  The calculated probability distribution is convoluted with a Lorentzian
  of 200~MHz width.
}
\label{fig:Stark_diagram_standin}
\end{figure}

Figure~\ref{fig:SFI_spectra} presents state-selective field ionization (SFI)
spectra recorded at $n=60$.  Earlier studies have shown that the field at
which a given state ionizes is governed by $n$, the magnitude of the
magnetic quantum number $\vert M_{L}\vert$, and the
slew rate of the applied field~\cite{sd83,g94,fhv15}.  Briefly, for a hydrogen
atom, the ionization threshold depends on the orientation of the atomic dipole 
moment which is defined by the electric quantum number, $k$.
The extreme red-shifted states within a given Stark $n$ manifold for which $k\sim -n$
ionize at the smallest fields, given by $\sim1/(9 n^4)$, because their electronic
wave functions are polarized towards the saddle point in the electron
potential that results from application of the external field.  
Stark states whose electronic wave functions are concentrated perpendicular to the ionization field ($k\sim0$) and blue-shifted Stark states ($k\sim n$) which are polarized away from the saddle point both have higher energies but, despite this, their field ionization thresholds, $\sim 1/(6n^4)$ and $\sim 1/(4n^4)$, respectively, are higher due to the reduced electron probability density in the vicinity of the saddle point. For non-hydrogenic atoms,
short-range interactions between the Rydberg electron and the core electrons
lead to dramatic changes in the ionization characteristics. 
Core penetration leads to sizable quantum defects
for the non-degenerate low-$L$ states and to the appearance of avoided crossings between Stark states.  Such avoided crossings are evident in the calculated Stark diagram for strontium shown in Figs.~\ref{fig:Stark_diagram}\,(a) and (b). 
In these calculations the wave functions for low-$L$ ($L\leq 6$) states are
obtained using a two-active-electron model and the higher $L$ states are
approximated by hydrogenic wave functions \cite{PhysRevA.90.013401}.  
Classically, the energy separations at the avoided crossings represent the precession
rates associated with scattering of the Rydberg electron by the core ion.
When the slew rate of the ionizing field is small enough,  
even the blue-shifted Stark states can precess sufficiently rapidly to follow the changes in the electron wave function that accompany the traversal of an avoided crossing. 
The avoided crossings are therefore traversed adiabatically as the field is increased and ionization
occurs when the electronic energy lies above the saddle point 
at fields $\sim1/(16 n^4)$.
For sufficiently high slew rates, such adiabatic passage is no longer possible
and the avoided crossings are traversed diabatically, thereby mimicking
the ionization dynamics of a hydrogen atom for which the threshold field
depends on the initial polarization of the electronic wave function.
Similarly,  with increasing $\vert M_{L}\vert$ (the projection of $L$ onto the 
axis of the ionization field), the effect of core electron scattering is 
suppressed due to the centrifugal barrier and the ionization
behavior again becomes similar to that of a hydrogen atom.

An example of largely adiabatic behavior is evident in Fig.~\ref{fig:Stark_diagram}\,(a) 
which shows the evolution of the electron probability distribution 
for $n=60$ $ ^{3}\text S _{1} (M_{J}=1)$ atoms during SFI calculated using an electric
field slew rate of 43 V cm$^{-1}\mu$s$^{-1}$, similar to that used
experimentally. The bulk of the crossings are traversed adiabatically 
resulting in little change in the electronic energy as the field 
is increased.  (This behavior can be also seen in 
Fig.~\ref{fig:Stark_diagram_standin}.) 
Eventually ionization occurs at a field of
$\sim1/16(n-\delta)^{4}$, which for $n=60$ is $\sim 32 $ V cm$^{-1}$.
With the laser tuned to atomic 
resonance, field ionization results in a narrow feature centered 
near 35~V~cm$^{-1}$ (see Fig.~\ref{fig:SFI_spectra}).  Given the uncertainty in the applied field, the data therefore
suggest that ionization of parent $60s$ atoms occurs along principally 
adiabatic paths. 
As shown in Fig.~\ref{fig:Stark_diagram_standin}, similar largely adiabatic ionization is predicted for the neighboring 
$n=59$ $ ^{3}$D$_{3}$ ($M_{J}=3$) state. 
As $\vert M_{L}\vert$ is increased, 
an increasing fraction of the avoided crossings are traversed diabatically, 
leading to ionization along principally diabatic paths.  This is illustrated in
Fig.~\ref{fig:Stark_diagram_standin} by the ionization trajectories 
calculated for the $n=56$ $^{3}$G$_{5}$ ($M_{J}$=5) state.
$G$ states have a small but finite quantum defect
shifting the energy to the red compared to the quasi-degenerate higher-$L$ states.
Therefore, upon application of the SFI field, they follow the ionization paths
of the most red-shifted Stark states.

Detuning the laser to create Rydberg molecules leads to marked changes in the
SFI profile.  A single SFI feature is still observed whose threshold is
similar to that for the parent atomic state but that is broadened towards
higher fields and is rather asymmetric. Since the SFI profiles in
Fig.~\ref{fig:SFI_spectra} were recorded within $1-2.5 \mu$s from the end of
the laser pulse and are insensitive to the exact time delay, it is reasonable
to conclude that any effects of collisions and of blackbody-radiation-driven
transitions must, at least on this time scale, be small. This suggests that the differences in
the SFI profiles seen for the atomic and molecular Rydberg states result primarily from the presence of ground state atoms within the electron orbit and that these atoms perturb the energy separations at avoided crossings resulting in a more diabatic path to ionization.  (Similar, although less pronounced, broadening was observed in earlier studies of lower-$n$ strontium Rydberg molecules formed in less-dense thermal samples \cite{PhysRevA.93.022702}.)

The presence of ground state atoms strongly modifies the energy level structure that is associated with Rydberg molecules. Due to the coupling of
the electronic motion with the vibrational and rotational degrees of
freedom of the molecule, a single atomic electronic state is split into multiple rovibronic levels. 
Typical rovibronic states of the strontium Rydberg dimer are shown in
Figs.~\ref{fig:Stark_diagram}(c) and \ref{fig:Stark_diagram}(d).
When the atomic Stark states are well isolated (such as indicated
by the points labeled c and d in Fig.~\ref{fig:Stark_diagram}(b)), the associated
molecular levels can be evaluated by first-order perturbation theory 
using a Fermi pseudopotential
\begin{equation}
V(\vec{r})=2\pi A_{S}(k)\delta(\vec{r}-\vec{R}_g)
\end{equation}
where $\vec{r}$ and $\vec{R}_g$ are the positions of the Rydberg electron 
and ground-state atom.  (We
note that, since the present analysis is largely qualitative, the $p$-wave
scattering length is neglected for simplicity.)   
Figures~\ref{fig:Stark_diagram}(c) and \ref{fig:Stark_diagram}(d) display cross sections of the molecular potential 
taken at $\vec{R}_g = (0, 0, z)$ and of the corresponding 
molecular wavefunctions for the lowest lying 16 rovibronic levels. The wave functions are
shifted by the binding energy $E_\lambda$ and normalized by a factor $C$
so that each wave function, $C | R_g \psi_\lambda(\vec{R}_g) |^2$, can be distinguished.
At the avoided crossings between the atomic Stark states 
the molecular rovibronic states also form avoided crossings. However, 
their energy separations are much smaller due to the high density 
of rovibronic states. Therefore, for a given slew rate the likelihood of 
diabatic passage at avoided crossings is greater for molecules 
than for atoms. This would be true even if using a more accurate calculation that includes the contribution from p-wave scattering because, while this will lead to small shifts in the energies of rovibronic levels, there would be no dramatic changes in the density of states. Adiabatic traversal requires not only the precession of the
electronic wave function but also a deformation of the molecular wave function which, as suggested by the smaller
energy separations at avoided crossings, requires a much longer time scale. 
The energy separations are expected to be particularly small at crossings 
between red- and blue-shifted Stark states where the Franck-Condon overlap
is extremely small. 

\begin{figure}
\includegraphics[scale = 0.8]{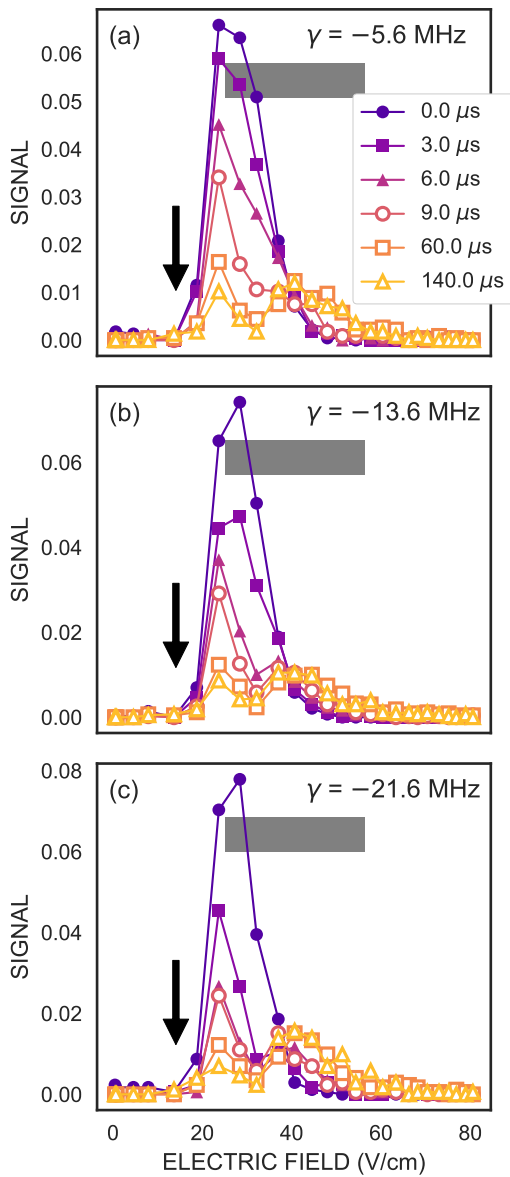}
\caption{Time evolution of the SFI signals recorded for $n=72$ and the detunings, $\gamma$, indicated.  The various time delays, $t_{D}$, are listed in the inset.  The arrows and gray bars denote the threshold field for adiabatic ionization and the range of fields encompassed by purely diabatic ionization, respectively.}
\label{fig:time_evolution}
\end{figure}

The additional complexity introduced by the presence of 
multiple ground-state atoms within the electron cloud can further increase the
probability of diabatic ionization.  Thus it is not surprising that the widths of the
molecular SFI profiles in Figs.~\ref{fig:SFI_spectra}(b)-(d) extend 
from $1/(16 (n-\delta)^4) \simeq 31$~V~cm$^{-1}$ up to 
$1/(9 (n-\delta)^4) \simeq 56$~V~cm$^{-1}$, i.e., from the adiabatic limit to
the diabatic limit of the most red-shifted Stark states with
which the $n=60$ $^{3}$S$_{1}$ state is most strongly coupled 
(Fig.~\ref{fig:Stark_diagram}(a)). Taking into account the rather poor statistics in Fig. \ref{fig:SFI_spectra} (b), the spectra are seen to be relatively insensitive 
to laser detuning, i.e., local ground-state atom
density, indicating that rovibronic
states play an important role in determining the field ionization behavior even when there are relatively few atoms within the electron cloud. 

Figure~\ref{fig:time_evolution} shows the time evolution of the SFI spectrum for $n=72$ molecules following the laser excitation pulse.  Even at early times the spectrum is quite broad and is peaked at fields substantially above the adiabatic threshold indicating that, for this value of $n$, a sizable fraction of the molecules follow largely diabatic paths to ionization. The measured profiles are, again, relatively insensitive to laser detuning.  The ionization signal at the lower ionizing fields decreases rapidly with increasing time delay, this being accompanied by a small increase in the ionization signal at the highest ionizing fields.  Taken together, these changes result in a marked change in the overall SFI profile. 

\section{ Lifetime of Rydberg molecules }

The time dependences of the SFI spectra are analyzed in this section 
by considering the collision
processes that can
lead to Rydberg molecule destruction~\cite{PhysRevX.6.031020}.  
At a temperature of 150~nK, and assuming an electric dipole polarizability of $C_4 = 186$ a.u. \cite{PhysRevA.10.1924}, the Langevin cross section, $\pi R_{\rm LG}^2$, for an ion-atom
collision is large, and corresponds to a value $R_{\rm LG} \simeq 280$~nm which is 
comparable to the radius of an $n=53$ $^3S_1$ Rydberg atom.  Given that the present atomic densities correspond to most probable nearest-neighbor atomic separations of $\sim100$~nm which are significantly less than $R_{LG}$, and that the size of the present Rydberg molecules is such that at such radii screening of the core ion by the Rydberg electron is incomplete, it is reasonable to view collisions in terms of a binary interaction between the core ion and nearest ground-state atom.  At small separations, such interactions
are dominated by the Sr$^{+}$-Sr potential which correlates with two electronic
states, the gerade (bonding) V$^+_g$ and ungerade (antibonding) V$^+_u$
states. Quasi-resonant energy transfer between the Rydberg electron and
these molecular ionic states can lead to inelastic processes~\cite{PhysRevA.20.1890,PhysRevA.21.819,PhysRevX.6.031020}. 
The first process is associative ionization
\begin{equation}\label{eqn:associative ionization}
\text S\text r^{\text M}(n\,^{3} \text S _{1}) +\text S \text r \rightarrow \text S\text r _{2}^{+} + \text e ^{-} ,
\end{equation}
where $\text S\text r^{\text M}$ denotes the Rydberg atom forming the core of a
molecule. This inelastic collision is one of a family of a so-called chemi-ionization
reactions which lead to ionization of one of the collision partners or to
formation of new chemical bonds, i.e., molecular ions. 
Upon ionization the excess energy is carried off by the Rydberg electron. 
Another inelastic process that can occur is $L$-changing 
\begin{equation}\label{eqn:L-changing}
\text S\text r^{\text M}(n\,^{3} \text S _{1}) +\text S \text r \rightarrow \text S\text r (n'L') + \text S\text r 
\end{equation}
in which the electron transitions to a nearby lower-lying Rydberg state.  The energy released, $\sim1/n^{3}$, ($\sim30$~GHz at $n=60$), is much greater than the molecular binding energy and is sufficient to give the core ion and the ground-state atom with which it is interacting velocities of $\gtrsim10$~ms$^{-1}$.  The excited electron remains bound to the core ion producing a fast Rydberg atom that escapes the ODT on a time scale of $\sim1\mu$s after which it undergoes no further collisions.  As will be discussed, $L$-changing collisions mainly populate low-$L$ states.  The $L$-changed Rydberg atoms, however, remain within one or
two millimeters of the ODT over the duration of the present experiments and
thus contribute to the total SFI signal.  The broadening of the SFI
profiles seen at late times in
Fig.~\ref{fig:time_evolution} provides evidence of
$L$-changing reactions.  Similar changes in the SFI profiles were also seen for $n=49$
and $n=60$. 

\begin{figure}
\includegraphics[scale=0.8]{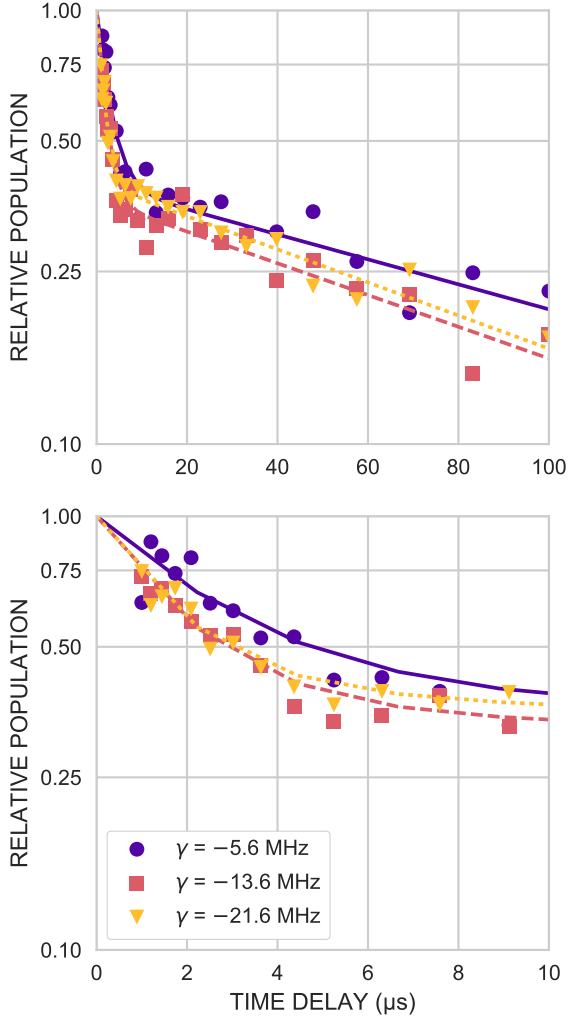}
\caption{Top: Time dependence of the total SFI signal for $n=49$ and the detunings, $\gamma$, indicated. The data sets are each normalized to one at $t=0$.  The solid, dashed and dotted lines show fits to the data obtained using Eqs.~\ref{lpop} for detunings of $-5.6$ MHz, $-13.6$ MHz and $-21.6$ MHz respectively (see text). Bottom: Expanded view at early delay times.}
\label{fig:total_SFI}
\end{figure}

Figure~\ref{fig:total_SFI} shows the time evolution of the total SFI signal for $n=49$ and the detunings indicated.   (Similar time evolutions are seen for $ n = 60$ and 72.) For each detuning the total SFI signal initially decreases rapidly, the rate of decrease slowing markedly at late times. Given that the products of $L$-changing reactions are included in the total SFI signal, the initial rapid rate of loss must result from associative ionization. At late times the bulk of the SFI signal is associated with $L$-changed atoms which have been ejected from the ODT. Once ejected these atoms undergo no further collisions and decay through radiative processes, i.e., spontaneous emission or interactions with blackbody radiation. 

\begin{figure}
\includegraphics[scale=0.8]{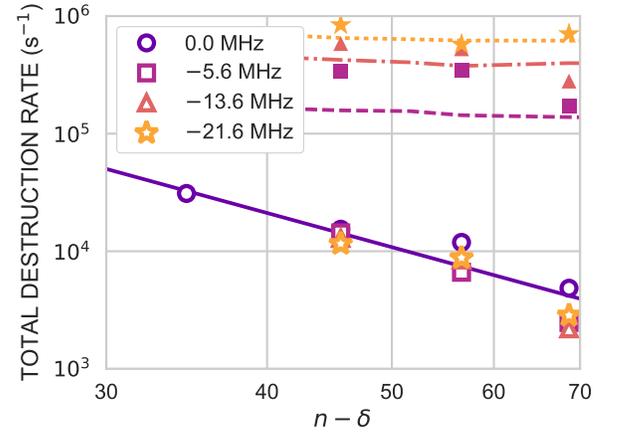}
\caption{Closed symbols: Rydberg molecule destruction rates, $\Gamma_{LOSS}=(\Gamma_{AI}+\Gamma_L)$,
  measured using the detunings, $\gamma$,
  indicated. The dashed, dot dashed and dotted lines show the results of model simulations for detunings of $-5.6$ MHz, $-13.6$ MHz and $-21.6$ MHz respectively (see text). Open symbols: decay rates, $\Gamma_R$, measured using thermal
  samples on atomic resonance \cite{PhysRevA.93.022702} for zero detuning 
  and $L$-changed atoms created in the BEC for finite detunings. The solid line indicates a fit to the decay rates measured on resonance.}
\label{fig:associative_ionization}
\end{figure}

\begin{figure}
\includegraphics[scale = .8]{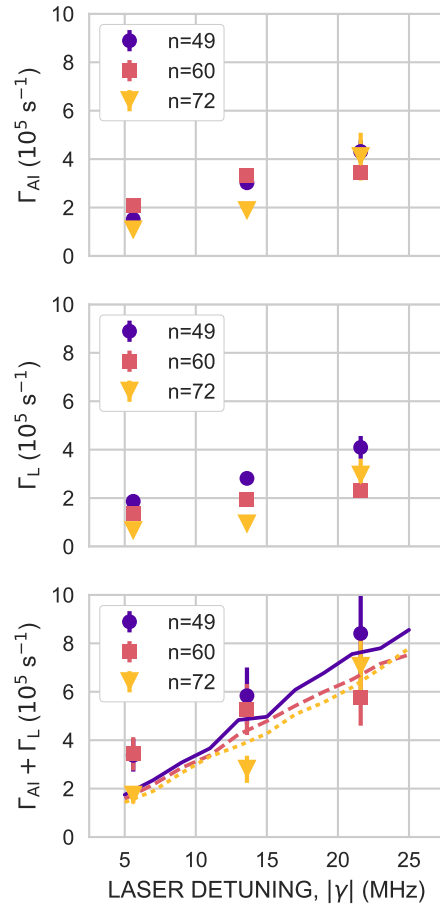}
\caption{Dependence of the reaction rates Top: $\Gamma_{AI}$, Middle: $\Gamma_L$,
  and Bottom: $\Gamma_{LOSS}=(\Gamma_{AI}+\Gamma_L)$ on detuning for the values of $n$ indicated. The results of model simulations are included using solid, dashed and dotted lines for principal quantum numbers of 49, 60 and 72 respectively.}
\label{fig:reaction rates}
\end{figure}

The experimental data can be fit using a reaction rate model which assumes that the parent
Rydberg molecules can be lost through associative ionization with rate
$\Gamma_{AI}$,  $L$-changing collisions with rate $\Gamma_L$, and radiative decay with rate $\Gamma_R$. The rate equation governing the population, $N_P$, of parent Rydberg molecules is thus
\begin{equation}
\frac{\mathrm{d} N_P}{\mathrm{d} t}=-(\Gamma_{AI}+\Gamma_L+\Gamma_{R})N_P.
\end{equation}
If it is further assumed that the radiative decay rate for the Rydberg
molecules equals that of the parent Rydberg atoms which, as will be shown, is also similar to
that of the $L$-changed Rydberg atoms, the time dependence of the $L$-changed Rydberg population,$N_L$, can be written as
\begin{equation}
\frac{\mathrm{d} N_L}{\mathrm{d} t}=\Gamma_L N_P-\Gamma_R N_L.
\end{equation}
(It is assumed that $L$-changed atoms undergo no further collisions as they exit the ODT.)
The analytic solutions of these equations
\begin{equation} \label{lpop}
\begin{split}
N_P&=N_0 e^{-(\Gamma_{AI}+\Gamma_L+\Gamma_{R} )t} \\
N_L&=N_0 \frac{\Gamma_L}{\Gamma_{AI}+\Gamma_{L}}e^{-\Gamma_R t} \left [ 1 - e^{-(\Gamma_{AI}+\Gamma_L )t} \right ]
\end{split}
\end{equation}
are then fit to the measured sum, $N_P+N_L$. The late time data, where $N_P$
is very small, are used first to obtain the radiative decay rate $\Gamma_R$. The complete measured profiles are then fit using Eqs.~\ref{lpop} and the adjustable parameters $N_0$, $\Gamma_{AI}$, and $\Gamma_L$. The resulting best fits (for $n=49$) are presented in Fig.~\ref{fig:total_SFI}.
 Rates $\Gamma_{AI}$ and $\Gamma_L$, and the total molecular destruction rate $\Gamma_{LOSS}$ ($=\Gamma_{AI} + \Gamma_L$), obtained in this manner are presented in Figs.~\ref{fig:associative_ionization} and~\ref{fig:reaction rates} and are sizable, $\sim 2-8\times 10^{5}$ s$^{-1}$. 

For a given $n$, the total destruction rate increases steadily with detuning, i.e., local atom density, whereas for a given detuning, the measured destruction rates are almost independent of $n$. Since, for a given detuning, the local ground-state atom density in the molecule is largely independent of $n$, (see Eq.~\ref{eq:large detunings}), the lack of a strong $n$-dependence in the destruction rate points to a loss process whose time scale is set by the distance from the Rydberg core ion to the nearest ground-state atom, i.e., to a loss process that depends on a close collision between the Sr$^{+}$ core ion and a neighboring ground-state atom. Increases in detuning, i.e., local atom density, lead to an approximately linear increase in the total destruction rate which is explained below.

The experimental observations are consistent with the predictions of a model which assumes that, because of the very low initial temperatures in the BEC, any initial relative motions of the Rydberg core ion and neighboring ground-state atoms must be very small.  Reaction is therefore presumed to be initiated by the mutual attraction between the core ion and nearest-neighbor ground-state atom.  Monte-Carlo sampling of initial conditions is used  to estimate the time required for these to collide, their mutual attraction being described by a C$_{4}/r^{4}$ potential \cite{PhysRevA.92.052515}.  Any effects of the Rydberg electron in screening the core ion-neutral interaction are neglected. In the simulations 1000 ground-state atoms are uniformly distributed within a box with the Sr$^{+}$ ion at the center. (Since 1000 atoms are always considered, the size of the box is varied to change the density.) The closest neutral atom to the Sr$^{+}$ ion core is then selected and the time they take to collide, given by
\begin{equation}
t=\int_{r_{\textup{initial}}}^{r_{\textup{final}}}\frac{\mathrm{d}r}{v (r)}= - \int_{r_{\textup{initial}}}^{r_{\textup{final}}}\frac{\mathrm{d}r}{\sqrt{\frac{2}{\mu}\left ( E_{\textup{coll}}+\frac{C_4}{r^4} \right )}}
\end{equation} 
where $r_{\textup{initial}}$ and $r_{\textup{final}}(=0)$ are the initial and
final inter-particle separations, respectively, $\mu$ is the reduced mass, and
$E_{\textup{coll}}$ is the initial collision energy, calculated (the negative sign is included to account for the fact that $r_{final} < r_{initial}$). Upon
collision, reaction, either associative ionization or $L$-changing, is
presumed to occur. The calculations are repeated for many initial ground-state atom distributions to obtain the cumulative distribution of collision times and, following the fitting procedures outlined in ref.\cite{PhysRevX.6.031020}, the mean collision time is extracted and with it the collisional loss rate.  Reaction rates obtained using this simple model are included in Figs.~\ref{fig:associative_ionization} and~\ref{fig:reaction rates}. The calculated rates are comparable to the measured destruction rates and increase linearly with local atom density, $\rho$. Such behavior results because $E_{\textup{coll}}$ is negligible relative to the interaction potential $C_4/r^4$ when the collision time may be written
\begin{equation}
t\propto-\int_{r_{\textup{initial}}}^{r_{\textup{final}}}r^2\mathrm{d}r\propto r _{initial}^3\,.
\end{equation}
The reaction rate, which is
proportional to $1/t$, thus scales as $1/r_{initial}^3$.  Since the average
nearest-neighbor distance is on the order of $\rho ^{-1/3}$, the reaction rate
will then scale linearly with $\rho$, mirroring the observed behavior. As seen in Fig.~\ref{fig:reaction rates}, the rates for Rydberg molecule destruction through associative ionization and through $L$-changing are similar, which parallels the behavior seen earlier with rubidium Rydberg molecules.

As is evident from Fig.~\ref{fig:associative_ionization}, the measured
radiative decay rates, $\Gamma_R$, for $L$-changed atoms are only slightly
less than those of the parent $n^{3}$S$_{1}$ Rydberg states measured on
resonance in a cold thermal gas.  This is at first sight surprising because
$L$-changing might be expected to lead to the population of a broad
distribution of high-$L$ states whose lifetimes are typically much longer than
those of low-$L$ states.  However, given that $L$-changing necessarily
involves a short-range interaction, the product states must have significant
electron probability densities near the core ion pointing to the population of
relatively low-$L$ states.  Rates for the loss of low-$L$
($L\leq6$), $n=60$ states through the combination of spontaneous emission and blackbody-radiation-induced decay were therefore calculated.  No dramatic decrease in the calculated loss rates with increasing $L$ was observed.  The calculated rates for $^{3}$G$_{J}$ and $^{3}$H$_{J}$ states range from $\sim$0.5 to $1.4\times10^{4}$ s$^{-1}$, consistent with the present observations.
As might be expected (see Fig.~\ref{fig:associative_ionization}), the
lifetimes of the $L$-changed atoms (which are rapidly ejected from the trap
and undergo no further collisions) are essentially independent of the
detuning, i.e., of the initial local atom density in the BEC. 
 
\section{Conclusions}
The present work shows that the presence of ground-state atoms within a Rydberg electron orbit can lead to marked changes in  field ionization behavior resulting in increasingly diabatic passage to ionization and a broadening of the SFI profile. The measurements also demonstrate that, even though no p-wave resonance is present, ultralong-range strontium Rydberg molecules excited in a dense BEC are rapidly destroyed, on time scales of a few microseconds, through associative ionization and $L$-changing reactions initiated by a simple ion-induced dipole interaction between the ``bare'' Rydberg core ion and nearest neighbor ground-state atom. Both loss processes, however, require the presence of the Rydberg electron near the interacting Rydberg core-ion - ground-state atom pair which suggests that, for very high $n$ values where the electron probability density near the core ion becomes small, the loss rates might begin to decrease. Since ion-induced dipole interactions are not restricted to strontium the present measured lifetimes place strict limits on the time scales over which studies involving Rydberg species in any cold dense atomic gas can be undertaken, and reduce the coherence times in such measurements. 

In the future it will be interesting to explore the properties of Rydberg molecules created in dense $^{87}$Sr samples where fermion statistics limit the probability for finding two atoms in close proximity  which should, in turn, significantly reduce collisional loss rates.

\begin{acknowledgments}
Research supported by the NSF under Grants Nos. 1205946 and 1600059, by the AFOSR under Grant No. FA9550-12-1-0267, the Robert A. Welch Foundation under Grants No. C-0734 and C-1844, by the FWF (Austria) under Grant No. P23359-N16, and the SFB Nextlite.  The Vienna Scientific Cluster was used in the calculations.
\end{acknowledgments}

\bibliography{bibliography_FC_lifetimepaper}

\end{document}